\documentstyle[12pt]{article}
\vspace{5mm}
\title{A NEW METHOD OF QUANTIZATION OF CLASSICAL SOLUTIONS}
\vspace{2mm}
\author{D.V.Antonov\\
Institute of Theoretical and
Experimental Physics, \\
117259 B.Cheremushkinskaya, 25, Moscow, Russia}
\date{}
\begin{document}
\maketitle

\newcommand{\be}{\begin{equation}}
\newcommand{\ee}{\end{equation}}

\vspace{10mm}
\centerline{\bf Abstract}

\vspace{5mm}
Using stochastic quantization method [1], we derive  equations for
correlators of quantum  fluctuations around the classical solution in the
massless $\phi^4$ theory. The obtained equations are then solved in the
lowest orders of perturbation theory, and the first correction to the free
propagator of a quantum fluctuation is calculated.

\vspace{3mm}
\section{Introduction}

\hspace{5mm} Recently new equations for vacuum correlators in the $\phi^3$
theory and in gluodynamics were derived via stochastic quantization method
[2].  These equations, alternative to the Schwinger--Dyson and
Makeenko--Migdal equations, allow one to connect correlators with different
number of fields, but, in contrast to the Schwinger--Dyson equations, they
are gauge--invariant in the case of gauge theories, and their mathematical
structure is simpler than the structure of Makeenko--Migdal equations.

In this letter we shall apply this approach to derivation of
a set of equations for correlators of quantum  fluctuations around the
classical solution (lipaton)  [3] in the massless $\lambda\phi^4$ theory.
To this end we solve the Langevin equation for quantum fluctuation through
Feynman--Schwinger path integral representation, and, introducing the
generating functional $Z[J]$, defined by the formula (4), apply to it
cumulant expansion [4,5]. The latter in this case has pure perturbative
meaning and corresponds to the usual semiclassical method of quantization of
classical solutions, suggested in [6] and developed in [7], providing
expansion in the powers of coupling constant (or in the powers
of Planck constant). The generating functional $Z[0]$, where one neglects
for simplicity the second term in the exponent in (3), has a meaning of a
one--loop expression for the effective potential, generated by quantum
fluctuations. This analogy becomes clear after applying to $Z[0]$  cumulant
expansion: an $n$--th term of cumulant expansion is an $n$--point  Green
function with $n$ lipatonic insertions.

After that we expand the obtained  system of bilocal approximation in the
two lowest orders of perturbation theory and solve the obtained equations in
the case, when the lipaton size is large enough (the particular meanings of
this approximation for each of the equations of bilocal approximation are
clarified in section 3). As a result we get the first constant correction to
the free propagator of a quantum fluctuation, which occurs to be of the
order of $\frac{1}{\rho^2}$, where $\rho$ is a lipaton size.

Note, that stochastic quantization method  was never applied before to
quantization of classical solutions.  We hope, that the suggested method
will be especially useful in gauge theories, bacause, as was shown in [2],
in this case it preserves gauge invariance. The corresponding equations for
quantum corrections to an instanton will be treated elsewhere.

The plan of the letter is the following. In section 2 we
derive a system of equations for correlators of quantum fluctuations in
bilocal approximation.
In section 3 we solve these equations up to the order  of
$\lambda^{1/2}$  under the assumption, that the lipaton size is
large enough, and compute the first correction to the free propagator of a
quantum fluctuation. The main results of the letter are summarized in the
Conclusion.

\vspace{3mm}
\section{Equations for quantum fluctuations in Gaussian approximation}

\hspace{5mm} We shall start with the action of Euclidean massless $\lambda
\phi^4$ theory, which has the form

$$ S = \int dx \left [ \frac{1}{2}(\partial_{\mu}\phi)^2 -
\frac{\lambda}{4}\phi^4 \right ],~~~~~~ \lambda > 0. $$
In this theory exists a classical solution (lipaton)

$$\phi_{cl}(x) =
2\sqrt{\frac{2}{\lambda}}~\frac{\rho}{(x-x_0)^2+\rho^2}$$
with a centre $x_0$ and of an arbitrary size $\rho$ [3].

Let the lipaton with the centre $x_0=0$ be a classical part of the total
field $\phi: \phi=\phi_{cl}+\varphi$, where $\varphi$ is a quantum
fluctuation in the lipatonic background. This splitting yields the following
Langevin equation for $\varphi$:

$$\dot{\varphi} - (\partial^2 + 3\lambda\phi^2_{cl}+
3\lambda\phi_{cl}\varphi +
\lambda\varphi^2)\varphi = \eta,$$
where

\be
<\eta(x,t)\eta(x',t')> = 2\delta(x-x')\delta(t-t'),
\ee

$$<...> \equiv \frac{\int D\eta(...)exp\left [ -\frac{1}{4}\int
dx~dt~\eta^2(x,t) \right ]}{\int D\eta ~exp \left [ - \frac{1}{4}\int dx~dt
~\eta^2(x,t)\right ]}.$$

Considering Langevin time as a proper time, one can solve this equation via
Feynman--Schwinger path integral representation. The solution,
corresponding to the retarded Green function, has the form:

\be
\varphi(x,t) = \int\limits^t_0 dt' \int dy (Dz)_{xy} K_z(t,t')
F_z(t,t')\eta(y,t'),
\ee
where

$$K_z(t,t') \equiv \theta(t-t')exp \left \{ \int\limits^t_{t'} d\xi \left [
-\frac{\dot{z}^2(\xi)}{4} + 3\lambda\phi^2_{cl}(z(\xi)) \right ] \right
\},~~~~~(Dz)_{xy} = \lim_{N\rightarrow \infty}
\prod\limits^N_{n=1}\frac{d^4z(n)}{(4\pi\varepsilon)^2},$$

$$N\varepsilon = t-t', ~~~~z(\xi=t')=y, ~~~~z(\xi=t)=x $$

\be
\mbox{and}~~~ F_z(t,t') \equiv exp\left \{ \lambda \int\limits^t_{t'} d\xi \left [
3\phi_{cl}(z(\xi))\varphi(z(\xi),\xi)+\varphi^2(z(\xi),\xi) \right ] \right
\}.
\ee

In order to obtain a minimal closed set of equations for correlators
$<\varphi>,~ <\varphi\eta>,$\\
$<\varphi\varphi>,$ we introduce the
generating functional [2]:

\be
Z[J] = <F_{z}(t,t')F_{\bar{z}}(\bar{t},\bar{t}') exp\Biggl [\int
dud\tau J(u,\tau)\eta(u,\tau) \Biggr ] >
\ee
and apply to it cumulant
expansion [4,5]. Differentiating it twice by $J$, putting then $J=0$
and assuming, that the total  stochastic ensemble of fields $\varphi$
and $\eta$ is Gaussian, one gets the following system of equations of
bilocal approximation:

$$<\varphi(x,t)> = 3 \lambda\int\limits^t_0 dt'\int
dy(Dz)_{xy}K_z(t,t')\int\limits^t_{t'}d\xi~\phi_{cl}(z(\xi))\cdot $$

\be
\cdot <\varphi(z(\xi),
\xi)\eta(y,t')><F_z(t,t')>,
\ee

$$
<\varphi(x,t)\eta(\bar{x},\bar{t})> =
2\int(Dz)_{x\bar{x}}K_z(t,\bar{t})<F_z(t,\bar{t})> +
9\lambda^2\int\limits^t_0 dt'(Dz)_{xy}K_z(t,t') \cdot$$

\be
\cdot \int\limits^t_{t'}d\xi\int\limits^t_{t'}d\xi' \phi_{cl}(z(\xi))
\phi_{cl}(z(\xi'))<\varphi(z(\xi),\xi)\eta(y,t')><\varphi(z(\xi'),\xi')
\eta(\bar{x},\bar{t})><F_z(t,t')>,
\ee

$$<\varphi(x,t) \varphi(\bar{x},\bar{t})> =
2\int\limits^{min(t,\bar{t})}_0 dt'\int dy(Dz)_{xy} (D\bar{z})_{\bar{x}y}
K_z(t,t')K_{\bar{z}}(\bar{t},t')<F_z(t,t')F_{\bar{z}}(\bar{t},t')> +$$

$$+ 9\lambda^2 \int\limits^t_0 dt' \int\limits^{\bar{t}}_0 d\bar{t}' \int
dy~d\bar{y}(Dz)_{xy}(D\bar{z})_{\bar{x}\bar{y}}K_z(t,t')K_{\bar{z}}
(\bar{t},\bar{t}')\Biggl [ \int\limits^t_{t'}d\xi
\phi_{cl}(z(\xi))<\varphi(z(\xi),\xi)\eta(y,t')>+ $$

$$+ \int\limits^{\bar{t}}_{\bar{t}'}
d\xi'\phi_{cl}(\bar{z}(\xi'))<\varphi(\bar{z}(\xi'),\xi')\eta(y,t')>
\Biggr ] \Biggl [ \int\limits^t_{t'}
d\xi\phi_{cl}(z(\xi))<\varphi(z(\xi),\xi)\eta(\bar{y},\bar{t}')> + $$

\be
+\int\limits^{\bar{t}}_{\bar{t}'}d\xi'\phi_{cl}(\bar{z}(\xi'))<\varphi(\bar{z}
(\xi'),\xi') \eta(\bar{y},\bar{t}')> \Biggr ]
<F_z(t,t')F_{\bar{z}}(\bar{t},\bar{t}')>,
\ee
where

$$<F_z(t,t')> = exp\left \{\lambda\int\limits^t_{t'}d\xi \Biggl
[ 3\phi_{cl}(z(\xi))<\varphi(z(\xi),\xi)> +
<\varphi^2(z(\xi),\xi)> + \right.$$

\be
\left.+ \frac{9\lambda}{2}\int\limits^t_{t'}d\xi'\phi_{cl}(z(\xi))\phi_{cl}
(z(\xi'))
 \Biggl ( <\varphi(z(\xi),\xi)\varphi(z(\xi'),\xi')> -
<\varphi(z(\xi),\xi)><\varphi(z(\xi'),\xi')> \Biggr ) \Biggr ] \right \},
\ee

$$<F_z(t,t')F_{\bar{z}}(\bar{t},\bar{t}')> = exp\left
\{\lambda\int\limits^t_{t'} d\xi \Biggl [
3\phi_{cl}(z(\xi))<\varphi(z(\xi),\xi)> + <\varphi^2(z(\xi),\xi)> \Biggr
] \right. +$$

$$+\lambda\int\limits^{\bar{t}}_{\bar{t}'} d\xi' \Biggl [
3\phi_{cl}(\bar{z}(\xi'))<\varphi(\bar{z}(\xi'),\xi')> + <
\varphi^2(\bar{z}(\xi'),\xi')> \Biggr ] + \frac{9\lambda^2}{2}\Biggl [
\int\limits^t_{t'} d\xi \int\limits^t_{t'} d\xi'
\phi_{cl}(z(\xi))\phi_{cl}(z(\xi')) \cdot$$

$$\cdot < \varphi(z(\xi),\xi)\varphi(z(\xi'),\xi')>+
\int\limits^{\bar{t}}_{\bar{t}'} d\bar{\xi} \int\limits^{\bar{t}}_{\bar{t}'}
d\bar{\xi}' \phi_{cl}(\bar{z}(\bar{\xi})) \phi_{cl}(\bar{z}(\bar{\xi}'))
<\varphi(\bar{z}(\bar{\xi}),\bar{\xi})
\varphi(\bar{z}(\bar{\xi}'),\bar{\xi}')> +$$

$$+ 2 \int\limits^t_{t'} d\xi \int\limits^{\bar{t}}_{\bar{t}'}
d\bar{\xi} ~\phi_{cl}(z(\xi)) \phi_{cl} (\bar{z}(\bar{\xi}))
<\varphi(z(\xi),\xi) \varphi(\bar{z}(\bar{\xi}),\bar{\xi})> - \Biggl (
\int\limits^t_{t'} d\xi~\phi_{cl}(z(\xi))<\varphi(z(\xi),\xi)> + $$

\be
\left.\left.\left.+ \int\limits^{\bar{t}}_{\bar{t}'} d\bar{\xi}\phi_{cl}
(\bar{z}(\bar{\xi})) <\varphi(\bar{z}(\bar{\xi}),\bar{\xi})> \right )^2
\right ] \right \}.
\ee
Note, that in the stochastic quantization
method the order of quantum correction of the grand ensemble of fields
$\varphi$ and $\eta$ is determined through the maximal number of the
Gaussian noise fields $\eta$ in correlators. Therefore, bilocal
approximation describes the first semiclassical approximation to the
classical lipatonic vacuum.

Solving this system of equations and  taking the asymptotics of the
equal--Langevin--time solutions, when Langevin time tends to infinity, one
obtains physical correlators $<\varphi>_{vac}$,\\
$<\varphi\varphi>_{vac}$.

\vspace{3mm}
\section{Solutions of equations (5)--(9) in the  lowest orders of
perturbation theory}

\hspace{5mm} Let us perform perturbative expansion of equations (5)--(9)
up to the order of $\lambda^{1/2}$. Expanding each of the
correlators in the powers of $\lambda^{1/2}$, one gets from
(5)--(9):

\be
<\varphi(x,t)>^{(0)} = <\varphi(x,t) \eta(\bar{x},\bar{t})>^{(1)} =
<\varphi(x,t)\varphi(\bar{x},\bar{t})>^{(1)} = 0,
\ee

\be
<\varphi(x,t) \eta(\bar{x},\bar{t})>^{(0)} = 2\int(Dz)_{x\bar{x}}
K_z(t,\bar{t}),
\ee

\be
<\varphi(x,t)\varphi(\bar{x},\bar{t})>^{(0)} =
2\int\limits^{min(t,\bar{t})}_0  dt' \int dy(Dz)_{xy}(D\bar{z})_{\bar{x}y}
K_z(t,t') K_{\bar{z}}(\bar{t},t'),
\ee

\be
<\varphi(x,t)>^{(1)} = 12\sqrt{2} \int\limits^t_0 dt' \int\limits^t_{t'}
d\tau \int dy (Dz)_{xy}K_z(t,t')\frac{\rho}{z^2(\tau)+\rho^2} \int
(Dz')_{z(\tau)y} K_{z'}(\tau,t').
\ee

In order to compute the integrals in (11)--(13), we make an assumption,
that the lipaton size $\rho$ is large enough.
In the case of equation (11), one, taking into account, that the integral in
the right hand side of this equation is saturated near the classical
trajectory, may write this condition  in the form $\rho\gg \mid x \mid +
\mid \bar{x}\mid$, so that all the terms of the  order of
$\underline{\underline{O}}(\frac{z^2(\xi)}{\rho^2})$  may be neglected in
comparison with 1. Then equation (11) yields in the physical limit,when
$T\equiv t - \bar{t}$ tends to infinity:

\be
\frac{\omega^2}{2\pi^2}
exp  \Biggl [ (\sqrt{3}-2)\omega T - \frac{\omega}{4}(x^2+\bar{x}^2) \Biggr
],
\ee
where $\omega \equiv \frac{8\sqrt{3}}{\rho^2}$.

Note, that this correlator is a purely lipatonic effect, since it
tends to zero, when the lipaton infinitely grows, so that its influence
becomes negligible.  

In the case of the integrals in the right hand side of equation (12) the
approximation of a large lipaton means, that we cut the region of
integration over $y$ by the following conditions:

\be
\mid y \mid \ll \rho - \mid x \mid,~~~~~~~ \mid y \mid \ll \rho - \mid
\bar{x} \mid.
\ee
We shall show below, that the contribution to $ <
\varphi(x,t)\varphi(\bar{x},t)>^{(0)}$ from the integration over all the
other values of $y$ in the case, when

\be
\mid x \mid + \mid \bar{x} \mid \ll \rho,
\ee
is much smaller than the contribution from the integration over the region
(15), which is equal to

$$<\varphi(x,t)\varphi(\bar{x},t)>^{(0)}=
\frac{\omega^2}{8\pi^2}\int\limits^t_0\frac{d\tau}{sh^2(2\omega\tau)}exp
\left \{ \frac{\omega}{4sh(2\omega\tau)} \Biggl [ 2x\bar{x} - \right.$$

\be
\left.-(x^2 + \bar{x}^2) ch(2\omega\tau) \Biggr ] + 2\sqrt{3}~\omega \tau
\right \}.
\ee
It is easy to obtain from (17), that, when the lipaton is infinitely large,
i.e. when $\omega \rightarrow 0$,

$$<\varphi(x,t)\varphi(\bar{x},t)>^{(0)}\longrightarrow
\frac{1}{4\pi^2(x-\bar{x})^2}e^{-\frac{(x-\bar{x}^2)}{8t}},$$
that in the limit $t \rightarrow +\infty$  yields an ordinary free boson
propagator. This fact indirectly indicates, that the contribution of the
region (15) is dominant.

In the case (16), splitting the interval of integration into two parts, $0
\leq \tau \leq \frac{1}{2\omega}$ and $\frac{1}{2\omega} \leq \tau \leq t$,
and using in every interval the corresponding asymptotics of the integrand,
one gets

$$<\varphi(x,t)\varphi(\bar{x},t)>^{(0)} = -\frac{3^{1/4}i}{8\pi}
\frac{\sqrt{\omega}}{\mid x - \bar{x}\mid} H^{(2)}_1 (3^{1/4}
\sqrt{\omega} \mid x -  \bar{x}\mid) +$$

\be
+ \frac{2+\sqrt{3}}{4\pi^2} \omega e^{-\frac{\omega}{4}(x^2+\bar{x}^2)}
 (e^{\sqrt{3}-2} - e^{2(\sqrt{3}-2)\omega t}),
\ee
where $H^{(2)}_1$ is the Hankel type--2  function. Taking the asymptotics of
(18) due to (16), we have

\be
\lim_{t\to +\infty}<\varphi(x,t)\varphi(\bar{x},t)>^{(0)}
= \frac{1}{4\pi^2(x-\bar{x})^2} + \frac{(2+\sqrt{3})e^{\sqrt{3}-2}}
{4\pi^2}\omega.
\ee
 The contribution to $<\varphi(x,t)\varphi(\bar{x},t)>^{(0)}$ from the
 integration over those values of $y$, which do not satisfy the condition
 (15) (i.e. from the region, where the  influence of the lipaton is
 negligible, and the motion is approximately free), is equal to
 $\frac{\omega}{64\sqrt{3}\pi^2}$. It is of the same order as the second
 term in the right hand side of (19) and, due to (16), is much smaller than
 the first term. Hence, the first correction to the free propagator of
 quantum fluctuation is a constant, which is equal to

\be
\frac{\omega}{4\pi^2}\Biggl ( (2+\sqrt{3})e^{\sqrt{3}-2} +
\frac{1}{16\sqrt{3}} \Biggr ).
 \ee

Finally, using the approximation of the large lipaton in the region
$\mid y \mid \ll \rho - \mid x \mid$, one gets from equation (13) in the
limit $t \rightarrow +\infty$:

\be
<\varphi(x,t)>^{(1)} =
\frac{3^{3/4}\omega^{5/2}e^{(\sqrt{3}-2)\omega t
- \frac{\omega x^2}{4}}}{32\pi^2} \int\limits^t_0 dt'\int\limits^t_{t'} d\tau
a^2 \left \{ 1+ \frac{a-1}{\sqrt{3}[1+2cth(\omega(\tau-t'))]} \right \},
\ee
where

$$a \equiv \left \{ 1 - \frac{1}{4}sh(2\omega(\tau-t'))\Biggl [ 1 +
2cth(\omega(\tau-t'))\Biggr ] \right \}^{-1}.$$

We see, that $\lim\limits_{\omega\rightarrow 0}$$<\varphi(x,t)>^{(1)} = 0$,
because, when the influence of the lipaton is negligible, there can not
exist any corrections to $<\varphi(x,t)>^{(0)}$  of the order of
$\lambda^{1/2}$.

Note, that the integral in the right hand side of (21) diverges at $\tau =
t'$, since we used  unregularized  Langevin equation. In order to avoid this
divergence, one needs to smear\\
 $\delta(t-t')$  in (1), using stochastic
regularization scheme, based on a non--Markovian genera-\\
lization of the
Parisi--Wu approach [8]. When $t$ tends to infinity, $<\varphi(x,t)>^{(1)}$,
defined through equation (21), vanishes as
$\underline{\underline{O}}(e^{(\sqrt{3}-2)\omega t})$.

In analogous way one may compute the leading term of the asymptotics of
$<\varphi(x,t)>^{(1)}$ at $t$ tending to infinity due to the region of
integration $\mid y \mid \geq \rho - \mid x \mid$, where the motion is
approximately free. It is equal to

\be
\frac{3^{5/4}}{64\pi^2\sqrt{\omega}t}\int\limits^t_0 dt'
\int\limits^t_{t'}\frac{d\tau}{(\tau-t')^2}
\ee
and also should be regularized as explained above. In the physical limit,
$t \rightarrow +\infty$, this expression tends to zero as
$\underline{\underline{O}}(\frac{1}{t})$.

\vspace{3mm}
\section{Conclusion}

\hspace{5mm} In this letter we
applied stochastic quantization to derivation of equations for vacuum
correlators of quantum corrections to the classical solution (lipaton)
in the massless $\lambda \phi^4$ theory in the approximation, that the grand
ensemble  of these quantum fluctuations and stochastic noise fields is
Gaussian.

The obtained equations are nonlinear integral equations of the second type
and contain functional integrals only in quantum mechanical sense as
integrals over trajectories, but not the integrals over fields, and,
therefore, may be easily investigated in the lowest orders of perturbation
theory. This work is performed in section 3 for the case, when the lipaton
size is large enough (the meaning of this approximation is clarified there).
The solutions are  given by the formulae (10),(14),(17)--(22), from which the
main one is the formula (20), which yields the first constant correction to
the free propagator of a quantum fluctuation, while the other correlators in
the order of $\lambda^{1/2}$, derived from the properly regularized
Langevin equation (see the end of section 3), vanish in the physical limit,
$t \rightarrow +\infty$, according to the formulae (14),(21)  and (22).

I am grateful to Professor Yu.A.Simonov for useful discussions and to
M.Markina for typing the manuscript.

The work is supported by the Russian Fundamental Research Foundation, Grant
No.93--02--14937.

\newpage
\noindent
{\bf References}

\vspace{5mm}

\noindent
1. G.Parisi and Y.Wu, Scienta Sinica {\bf 24}, 483 (1981);
for a review see

P.H.Damgaard and H.H\"uffel, Phys.Rep. {\bf 152}, 227-398 (1987).

\noindent
2. D.V.Antonov and Yu.A.Simonov,  International Journal

of Modern Physics A (in press).

\noindent
3. L.N.Lipatov, Pis'ma v ZhETF {\bf 25}, 116 (1977) (in Russian);
S.Fubuni,

Nuovo Cim. {\bf A34}, 521 (1976).

\noindent
4. N.G.Van Kampen, {\it Stochastic processes in physics and chemistry},

North-Holland Physics Publishing, Amsterdam, 1984; Yu.A.Simonov,

Yad.Fiz. {\bf 50}, 213 (1989) (in Russian).

\noindent
5. Yu.A.Simonov, Yad.Fiz. {\bf 54}, 192 (1991) (in Russian).

\noindent
6. R.Dashen et al., Phys.Rev. {\bf D10}, 4114, 4130, 4138 (1974); N.H.Christ

and T.D.Lee, Phys.Rev. {\bf D12}, 1606 (1975).

\noindent
7. C.Callan and D.Gross, Nucl.Phys. {\bf B93}, 29 (1975); J.Goldstone

and R.Jackiw, Phys.Rev. {\bf D11}, 1486 (1975); M.Creutz, Phys.Rev.
{\bf D12},

3126 (1975); E.Tomboulis, Phys.Rev. {\bf D12}, 1678 (1975); R.Rajaraman

and E.Weinberg, Phys.Rev. {\bf D11}, 2950 (1975); Yu.A.Simonov,

Yad.Fiz. {\bf 34}, 1640 (1981) (in Russian).

\noindent
8. J.D.Breit et al., Nucl.Phys. {\bf B233}, 61 (1984); J.Alfaro,
Nucl.Phys. {\bf B253},

464 (1985).
\end{document}